\documentclass[pra,amsmath,aps,twocolumn,superscriptaddress]{revtex4}
\usepackage{amsmath,mathrsfs,amsbsy,amssymb,lmodern,graphicx,bm,amsthm,amsfonts}
\usepackage{units}
\usepackage{bbm}
\usepackage{multirow,color}
\usepackage{subfigure}
\newcommand{\tr}{{\rm Tr}}
\begin{document}
\title{Genuine multipartite nonlocality in the one-dimensional ferromagnetic spin-1/2 chain}
\author{Yue Dai}
\affiliation{College of Physics, Optoelectronics and Energy, Soochow University, Suzhou, 215006, China}
\author{Chengjie Zhang}
\email{zhangchengjie@suda.edu.cn}
\affiliation{College of Physics, Optoelectronics and Energy, Soochow University, Suzhou, 215006, China}
\affiliation{Key Laboratory of Quantum Information, University of Science and Technology of China, CAS, Hefei, 230026, China}
\author{Wenlong You}
\affiliation{College of Physics, Optoelectronics and Energy, Soochow University, Suzhou, 215006, China}
\author{Yuli Dong}
\affiliation{College of Physics, Optoelectronics and Energy, Soochow University, Suzhou, 215006, China}
\author{C.H. Oh}
\affiliation{Centre for Quantum Technologies, National University of Singapore, 3 Science Drive 2, Singapore 117543, Singapore}
\affiliation{Physics Department, National University of Singapore, 3 Science Drive 2, Singapore 117543, Singapore}

\begin{abstract}
Genuine multipartite entanglement has been found in some spin chain systems. However, genuine multipartite nonlocality, which is much rarer than genuine multipartite entanglement, has never been found in any spin chain system. Here we present genuine multipartite nonlocality in a spin chain system. After introducing the definition of genuine multipartite nonlocality and a multipartite Bell-type inequality, we construct a group of joint measurements for the inequality in a one-dimensional ferromagnetic $N$-qubit chain with nearest-neighbor XXZ interaction, and many violations to the inequality have been found. The violations do indicate that genuine multipartite nonlocality exists in this ferromagnetic spin-1/2 chain system. Last but not least, we also calculate genuine multipartite entanglement concurrence in the same spin chain to demonstrate the difference and relationship between genuine multipartite nonlocality and genuine multipartite entanglement.
\end{abstract}

\maketitle

\section{Introduction}
Nonlocality is a salient quantum feature. Specifically, it denotes nonlocal correlation between two spatially separated systems. Nonlocality of quantum systems can be characterized via mathematical methods, one of which is called Bell inequality. It was first proposed by Bell in 1964 \cite{Bell}. Subsequently it was generalized to CHSH inequality \cite{CHSH,CGLMP}. These Bell-type inequalities provide a convenient way to test the local hidden variable theory. A system governed by local hidden variables should fulfill the inequalities. In other words, any violation of the inequalities can be interpreted as a sign of nonlocality. Recently, nonlocality has attracted much attention for its potential application in quantum computation \cite{qc1,qc2,qc3,qc4,qc5} and quantum communication \cite{qi}.


Compared with bipartite states, the structure of multipartite states is much more complex \cite{NE}. In a bipartite state, there has only one type of nonlocality. While in an $N$-partite multipartite state, there exist $N-1$ types of nonlocality. The growing number of types results in that there is a hierarchy in characterizing multipartite nonlocality \cite{hie}. Among these types of nonlocality, the first (or the weakest) type of multipartite nonlocality is a natural generalization of Bell's bipartite nonlocality \cite{MABK,WWZB,Hardy,Hardy2,Hardy3}. On the other hand, the last (or the strongest) type of multipartite nonlocality which is called genuine multipartite nonlocality (GMN) contains fully nonlocal correlations \cite{Svetlichny}. Similar to the Bell inequalities, varieties of inequalities have been developed to detect GMN in multipartite states, e.g., the Svetlichny inequality \cite{Svetlichny}, the generalized Svetlichny inequality \cite{multi,higher} and the extended Hardy-type inequality \cite{mh}.

It has been proved that genuine multipartite entanglement (GME) \cite{Entanglement} is equivalent to GMN for all pure multipartite entangled permutation symmetric states \cite{ns5}, such as GHZ states \cite{GHZ}, Dicke states \cite{Dicke}, and pure permutationally invariant multimode Gaussian states \cite{Gaussian}. This indicates a reliable method for searching nonlocal states. Preparing genuinely multipartite nonlocal states via quantum gates is difficult due to its fragility from decoherence. A more feasible access of obtaining nonlocal states is to produce the nondegenerate ground states (GSs) in a suitable system. In Ref. \cite{fs}, the authors pointed out that the GS of an anisotropic ferromagnetic spin$-1/2$ chain with length $N$ in a homogeneous magnetic field is a nondegenerate $N$-partite Dicke-like state. One can approximately get every type of the $N\!+\!1$ Dicke states from the GS by varying the external magnetic field and the inter-spin interaction strength.

In this paper, we start with the definition of GMN in the nonsignaling scenario and a multipartite Bell-type inequality for GMN. Then we construct a group of measurements according to the inequality for the one-dimensional $N$-qubit ferromagnetic chain with nearest-neighbor XXZ interactions, and calculate the result of the violation. It is found that the multipartite Bell-type inequality for GMN can be easily violated by the GSs of that anisotropic ferromagnetic spin$-1/2$ chain. Genuine multipartite entanglement has been found in some spin chain systems \cite{xy,ising}. But to our knowledge, there are few results about finding GMN in spin chain systems. Here we present GMN in the spin chain under the nonsignaling scenario. Furthermore, since all GMN states are GME states, the violation of the multipartite Bell-type inequality for GMN also indicates the existence of GME in the spin chain \cite{ns5}. Last but not least, we calculate GME concurrence in the same spin chain to demonstrate the difference and relationship between GMN and GME.

\begin{figure*}[htb]
\includegraphics[scale=0.67]{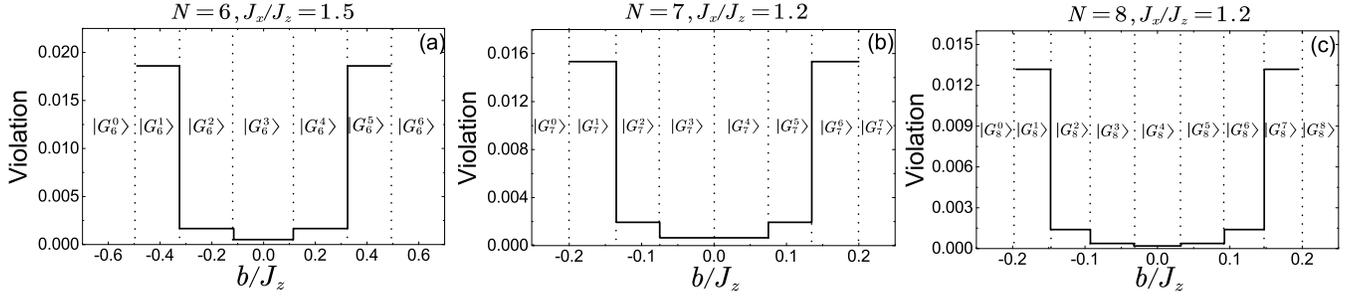}
\caption{The maximal violation of Eq. (\ref{chen}) is indicated by a solid line with changing $b/J_z$ and remaining $J_x/J_z$ unchanged. The value of the maximal violation of the $6,7,8$-partite spin chain are respectively shown in (a), (b) and (c). The maximum value of violation has a step-like plateaus and a reflection symmetry structure. Different storey denotes different $|G_{N}^{k}\rangle\ $ in subfigures. From left to right, the GS is $|G_{N}^{0}\rangle,|G_{N}^{1}\rangle,\cdots,|G_{N}^{N}\rangle$, respectively.}\label{fig1}
\end{figure*}

\section{Genuine multipartite Bell-type inequality}
Considering a system containing $n$ spacelike separated subsystems which are indexed with an array $I\!=\!\{1,2,\cdots,n\}$, the measurement setting and the outcome of the $k$-th $(k\!\in\!I)$ part of the system are recorded as $M_k$ and $r_k$, respectively. The joint probability distribution is denoted by $P(r_I|M_I)$, where $r_I=(r_1,\cdots,r_n)$ and $M_I=(M_1,\cdots,M_n)$. If the system belongs to a standard local realistic model, any nonlocal correlation cannot exist in such a system. Then the observed joint probability $P(r_I|M_I)$ satisfies
\begin{equation}\label{f1}
P(r_I|M_I)=\int \varrho(\lambda)\prod_{k=1}^{n}P_k(r_k|M_k,\lambda)\mathrm{d}\lambda,
\end{equation}
where $P_k(r_k|M_k,\lambda)$ is the probability of observer $k$ measuring observable $M_k$ with outcome $r_k$. The hidden variable $\lambda$ was distributed according to the normalized coefficient $\varrho(\lambda)$.

In a multipartite nonlocal correlation system, some qubits may share local correlations while others are nonlocal \cite{MABK,WWZB,Hardy,Hardy2,Hardy3}. GMN should also exclude such local-nonlocal models. The most general hybrid local-nonlocal model (or bi-local model) can be expressed as:
\begin{equation}\label{f2}
    P(r_I|M_I)=\sum_{\alpha}\int  \varrho_{\alpha}(\lambda)P_{\alpha}(r_{\alpha}|M_{\alpha},\lambda)P_{\bar{\alpha}}(r_{\bar{\alpha}}|M_{\bar{\alpha}},\lambda)\mathrm{d}\lambda,
\end{equation}
where $P_{\alpha}(r_{\alpha}|M_{\alpha},\lambda)$ is the joint probability distribution of every nonempty proper subset $\alpha=\{i_1,...,i_m\}$ of $I$ with a hidden variable $\lambda$ distributed according to $\varrho_{\alpha}(\lambda)$. The corresponding setting and the outcome are denoted by $M_{\alpha}=(M_{i_{1}},...,M_{i_{m}})$ and $r_{\alpha}=(r_{i_{1}},...,r_{i_{m}})$, respectively. $\bar{\alpha}$ is the complementary set of $\alpha$ in the set $I$. Note that $\bar{\alpha}=I\setminus\alpha$ is also a nonempty proper subset of $I$. In this hybrid local-nonlocal model, nonlocal correlations may be shared among all the subset of $\bar{\alpha}$ or $\alpha$, but not between $\bar{\alpha}$ and $\alpha$. This type of local-nonlocal model is called bi-locality.

In order to define GMN in the nonsignaling scenario, we have to suppose that all possible nonlocal correlations in $P_{\alpha}(r_{\alpha}|M_{\alpha},\lambda)$ and $P_{\bar{\alpha}}(r_{\bar{\alpha}}|M_{\bar{\alpha}},\lambda)$ are nonsignaling for any bipartite cut $\alpha$ and $\bar{\alpha}$ \cite{ns1,ns2,ns3,ns4} i.e., the marginal probability distribution of any part of the system is independent of the inputs on the remaining part. It is a natural condition since the allowing signaling is incongruous
with a physical perspective. This condition can be expressed as:

\begin{eqnarray}
&&P_{\beta\setminus \{k\}}\big(r_{\beta\setminus \{k\}}\mid M_{\beta\setminus \{k\}},\lambda\big)\nonumber\\
&=&\sum_{r_{k}} P_{\beta\setminus \{k\}}\big(r_{\beta\setminus \{k\}} r_k\mid M_{\beta\setminus \{k\}} M_k,\lambda\big)\nonumber\\
&=&\sum_{r_{k}} P_{\beta\setminus \{k\}}\big(r_{\beta\setminus \{k\}} r_k\mid M_{\beta\setminus \{k\}} M_k^{'},\lambda\big)\label{f3}
\end{eqnarray}
for all $k\in\beta$, and $\beta$ can be either $\alpha$ or $\bar{\alpha}$ with two or more
elements.

According to the above conditions, we give a strict definition of GMN in the nonsignaling scenario: A joint probability distribution $P(r_I|M_I)$ can be called GMN in the nonsignaling scenario if and only if it cannot be written in the form Eq. (\ref{f2}) with all possible nonlocal correlations being nonsignaling.

In an $n$-partite two-level system, the outcome $r_k$ of observer $k$ $(k\!\in\!{I})$ with measurements $M_k$ is confined to $\{0,1\}$, and $M_k$ has two alternative choices, e.g., $\{a_k,b_k\}$. For any $k,k'\!\in\!{I}$, we use notation $\bar{k}\!=\!I\setminus\{k\}$ and $\overline{kk'}\!=\!I\setminus\{k,k'\}$. One can obtain a multipartite Bell-type inequality according to the definition of GMN in the nonsignaling scenario \cite{ns5}:
\begin{eqnarray}\label{f4}
&&P(0_I|a_I)-\sum_{k\in I}P(0_I|b_k a_{\bar{k}})\nonumber\\
&&\ \ \ \ \ \ \  -\sum_{k\in I\setminus\{k'\}}P(1_{k'}1_k 0_{\overline{k'k}}|b_{k'}b_{k}a_{\overline{k'k}})\leq0,\label{chen}
\end{eqnarray}
where $P(r_I|M_I)$ should satisfy Eq. (\ref{f2}) and the nonsignaling condition Eq. (\ref{f3}). If the inequality (\ref{f4}) is violated, the $n$-partite system will be genuinely multipartite nonlocal. The detailed proof of the inequality (\ref{chen}) is given in Ref. \cite{ns5}.

\begin{figure*}[htb]
\includegraphics[scale=0.67]{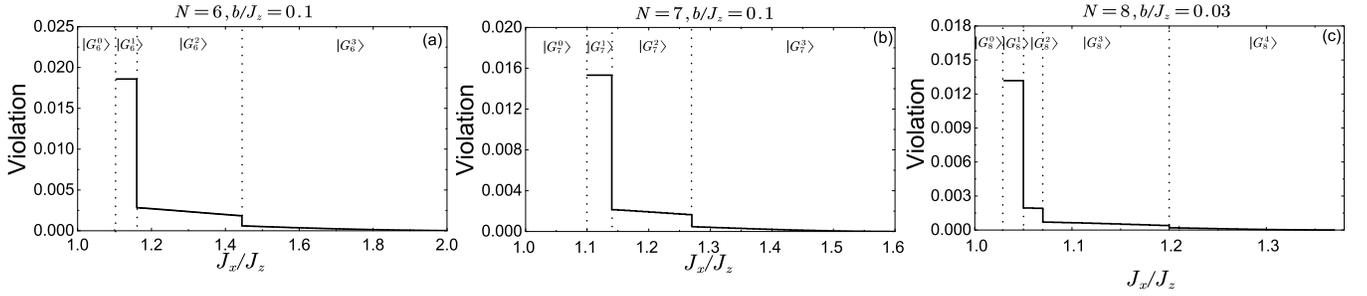}
\caption{The solid line indicates the maximal violation of Eq. (\ref{chen}) with changing $J_x/J_z$ and remaining $b/J_z$ unchanged.  The value of the maximal violation of the $6,7,8$-partite spin chain are respectively shown in (a), (b) and (c). In each subfigure, the GS is $|G_{N}^{0}\rangle,|G_{N}^{1}\rangle,\cdots,|G_{N}^{\lfloor N/2 \rfloor}\rangle$ respectively from left to right. The value of violation still varies gradiently at boundary values but decreases tardily with increasing of $J_x/J_z$.} \label{fig2}
\end{figure*}

\section{Genuine multipartite nonlocality in the one-dimensional ferromagnetic spin$-1/2$ chain}
Suppose that an anisotropic ferromagnetic spin$-1/2$ chain is placed in a homogenous magnetic field. The system can be described by the Hamiltonian:
\begin{equation}\label{f5}
\boldsymbol H\!=\!-J_{x}\!\sum_{j=1}^N (\boldsymbol S_{j}^{x} \boldsymbol S_{j+1}^{x}\!+\!\boldsymbol S_{j}^{y} \boldsymbol S_{j+1}^{y})\!-\!J_{z}\!\sum_{j=1}^N \boldsymbol S_{j}^{z} \boldsymbol S_{j+1}^{z}\!+\!b\!\sum_{j=1}^N \boldsymbol S_{j}^{z},
\end{equation}
where $\boldsymbol S_{j}$ is the $j$-th spin, ($j\!=\!{1,2,...N}$), and $\boldsymbol S_{N\!+\!1}\!=\! \boldsymbol S_1$. $J_x,J_z$ are the exchange coupling constants satisfying $J_x\!>\!J_z\!>\!0$. The coefficient $b$ denotes the strength of the external magnetic field \cite{fs2,fs3}. By varying $J_x/J_z$ and $b/J_z$, we obtain $N$-partite GS $\{|G_{N}^{k}\rangle\}_{k=0}^{N}$, and each $|G_{N}^{k}\rangle$ can be expressed as:
\begin{equation}\label{f6}
|G_{N}^{k}\rangle=\sum_{\alpha\subseteq{I},|\alpha|=k} \sqrt{P_{\alpha}}|0_{\bar{\alpha}}1_{\alpha}\rangle,
\end{equation}
where $\bar{\alpha}=I\setminus\alpha$ and $I=\{1,2,...,N\}$. $|\alpha|$ denotes the number of elements in  the subset $\alpha$ of $I$. $P_{\alpha}$ is the probability coefficient of the $N$-partite GS and $\sum_{\alpha\subseteq{I},|\alpha|=k} P_\alpha=1$. There exist total $\binom{N}{k}$ terms in $|G_{N}^{k}\rangle$. Notice that the GS is not equivalent to the standardized Dicke state $|D_{N}^{k}\rangle$:
\begin{equation}\label{f7}
|D_{N}^{k}\rangle=\binom{N}{k}^{-\frac{1}{2}} \sum_{\alpha\subseteq{I},|\alpha|=k} |0_{\bar{\alpha}}1_{\alpha}\rangle.
\end{equation}
All Dicke states should satisfy permutation symmetry so that all the probability $P_\alpha$ in Dicke states are equal, but generally speaking the probability $P_\alpha$ in GS (\ref{f6}) may be not equal.

Then we define a group of joint measurements on the $N$-partite ferromagnetic spin$-1/2$ chain according to Eq. (\ref{f4}):
\begin{eqnarray}\label{f8}
\hat{\mathscr{H}}&=& \hat{a}_{1}\hat{a}_{2}...\hat{a}_{n}- \hat{b}_{1}\hat{a}_{2}...\hat{a}_{n}- \hat{a}_{1}\hat{b}_{2}...\hat{a}_{n}-...\nonumber\\
&&- \hat{a}_{1}\hat{a}_{2}...\hat{b}_{n}- \hat{\overline{b}}_{1}\hat{\overline{b}}_{2}\hat{a}_{3}...\hat{a}_{n}
- \hat{\overline{b}}_{1}\hat{a}_{2}\hat{\overline{b}}_{3}...\hat{a}_{n}-...\nonumber\\
&&- \hat{\overline{b}}_{1}\hat{a}_{2}\hat{a}_{3}...\hat{\overline{b}}_{n}.
\end{eqnarray}
To simplify calculations, we set $\hat{a}_{1}\!=\!|a\rangle\langle a|$, $\hat{a}_{2},...,\hat{a}_{n}=|a'\rangle\langle a'|$, $\hat{b}_{1}=|b\rangle\langle b|$,  $\hat{b}_{2},...,\hat{b}_{n}=|b'\rangle\langle b'|$,  and choose a list of the form of every measurement:
\begin{eqnarray}\label{f9}
&&|a\rangle=\cos\theta_{1} |0\rangle+\sin\theta_{1} |1\rangle,\nonumber\\
&&|b\rangle=\cos\theta_{2} |0\rangle+\sin\theta_{2} |1\rangle,\nonumber\\
&&|\overline{b}\rangle=\sin\theta_{2} |0\rangle-\cos\theta_{2} |1\rangle,\nonumber\\
&&|a'\rangle=\cos\theta_{3} |0\rangle+\sin\theta_{3} |1\rangle,\nonumber\\
&&|b'\rangle=\cos\theta_{4} |0\rangle+\sin\theta_{4} |1\rangle,\nonumber\\
&&|\overline{b'}\rangle=\sin\theta_{4} |0\rangle-\cos\theta_{4} |1\rangle,\nonumber
\end{eqnarray}
where $\langle b|\overline{b}\rangle=0$, $\langle b'|\overline{b'}\rangle=0$, and $\theta_i\in {[0,\pi]}$ with $i=1,\cdots,4$. For each set of $J_{x}/J_{z}$ and $b/J_{z}$, the eigenvalues and eigenstates of the spin chain system can be calculated, and then we choose the eigenstate corresponding to the minimum eigenvalue as the GS. We can numerically get the maximum value of $\langle G_{N}^{k}|\hat{\mathscr{H}}|G_{N}^{k}\rangle$ with respect to $\{\theta_{i}\}$ after operating the above joint measurements on the GS $|G_{N}^{k}\rangle$. If $\langle\hat{\mathscr{H}}\rangle>0$, the inequality (\ref{f4}) is violated, and thus the GS is genuinely multipartite nonlocal. We have calculated the values of violation of $6,7,8$-partite ferromagnetic spin chain. The result is showed in Fig. \ref{fig1} and Fig. \ref{fig2}.

In Fig. \ref{fig1}, we change the parameter $b/J_z$ while the parameter $J_x/J_z$ remains fixed. In that case, the value of violation has a step-like plateaus. The magnetic field only determines the direction of the spin so that $b$ acting on the spin operator $S^{z}$ does not change eigenstates of the system. We denote the first two terms of Eq. (\ref{f5}) by
\begin{equation}
\boldsymbol H_{0}:=-J_{x}\!\sum_{j=1}^N (\boldsymbol S_{j}^{x} \boldsymbol S_{j+1}^{x}+\boldsymbol S_{j}^{y} \boldsymbol S_{j+1}^{y})\!-\!J_{z}\!\sum_{j=1}^N \boldsymbol S_{j}^{z} \boldsymbol S_{j+1}^{z},\nonumber
\end{equation}
and the last term by
\begin{equation}
\boldsymbol H_{1}:=b\!\sum_{j=1}^N \boldsymbol S_{j}^{z}.\nonumber
\end{equation}
One finds $[\boldsymbol H_{0}, \boldsymbol H_{1}]=0$. When $J_x/J_z$ keeps constant, varying $b/J_z$ only changes GS when it reaches a threshold. The GS will alter from one category of Dicke-like state to others. From left to right, the GS is $|G_{N}^{0}\rangle,|G_{N}^{1}\rangle,\cdots,|G_{N}^{N}\rangle$, respectively. The violation of $|G_{N}^{0}\rangle$ and $|G_{N}^{N}\rangle$ can not be calculated because $|G_{N}^{0}\rangle$ and $|G_{N}^{N}\rangle$ are not entangled states. The solid line in Fig. \ref{fig1}(b) seems to separate into five partitions but actually into six partitions. The structure of $|G_{N}^{k}\rangle$ is reversely symmetric to $|G_{N}^{N-k}\rangle$ so that two values of violation of $|G_{N}^{k}\rangle$ and $|G_{N}^{N-k}\rangle$ are equal.

In Fig. \ref{fig2}, we change the parameter $J_x/J_z$ while the parameter $b/J_z$ remains unchanged. From left to right, the GS is  $|G_{N}^{0}\rangle,|G_{N}^{1}\rangle,\cdots,|G_{N}^{\lfloor N/2 \rfloor}\rangle$ respectively, where the floor function $\lfloor x \rfloor$ is the greatest integer less than or equal to $x$. It is different from Fig. \ref{fig1} that the value of violation still varies gradiently at the threshold but decreases tardily with increasing of $J_x/J_z$.

\begin{figure}[tb]
\includegraphics[scale=0.7]{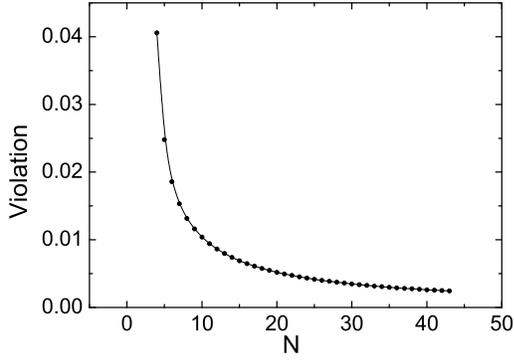}
\caption{Nodes indicate values of the maximal violation on $\langle G_{N}^{1}|\hat{\mathscr{H}}|G_{N}^{1}\rangle$ with each $N$ from $4$ to $43$. The analog function of the solid line suggests that the result approaches a limit value greater than zero.}\label{fig3}
\end{figure}

With the increase of the number $N$ of particles, the calculation becomes very complex. We only figure out the analytic solutions of the violation in the state $|G_{N}^{1}\rangle$. Then $J_x,J_z$ and $b$ should satisfy $J_z-J_x<b<(N-3)(J_z-J_x)/(N-1)$ \cite{fs}. The value of the violation in $|G_{N}^{1}\rangle$ can be written as:
\begin{eqnarray}\label{f11}
&&\langle G_{N}^{1}|\hat{\mathscr{H}}|G_{N}^{1}\rangle=\nonumber\\
&&\frac{1}{N} [\sin\theta_{1} \cos^{N-1}\theta_{3}+(N-1) \sin\theta_{3} \cos\theta_{1} \cos^{N-2}\theta_{3}]^2\nonumber\\
&&-\frac{1}{N}[\sin\theta_{2} \cos^{N-1}\theta_{3}+(N-1) \sin\theta_{3} \cos\theta_{2} \cos^{N-2}\theta_{3}]^2\nonumber\\
&&-\frac{N-1}{N}[\sin\theta_{1} \cos\theta_{4} \cos^{N-2}\theta_{3}+\sin\theta_{4} \cos\theta_{1} \cos^{N-2}\theta_{3}\nonumber\\
&&+(N-2)\sin\theta_{3} \cos\theta_{1} \cos\theta_{4} \cos^{N-3}\theta_{3}]^2\nonumber\\
&&-\frac{N-1}{N}[\sin\theta_{4} \cos\theta_{2} \cos^{N-2}\theta_{3}+\sin\theta_{2} \cos\theta_{4} \cos^{N-2}\theta_{3}\nonumber\\
&&-(N-2)\sin\theta_{2} \sin\theta_{3} \sin\theta_{4} \cos^{N-3}\theta_{3}]^2,
\end{eqnarray}
where $\hat{\mathscr{H}}$ is defined as Eq. (\ref{f8}). Then the maximal violation with $N$ from $4$ to $43$ has been calculated and shown in Fig. \ref{fig3}.

\section{Genuine multipartite entanglement in the one-dimensional ferromagnetic spin$-1/2$ chain}
In a bipartite system, entanglement means that the state $\varrho^{AB}$ cannot be written as $\sum_{k}p_{k}\varrho_{k}^{A}\otimes\varrho_{k}^{B}$. The status in multipartite systems is much more complex. To get started let us precisely define the underlying concepts of bi-separable and genuine multipartite entanglement.

An $N$-partite pure state $|\psi\rangle$ with Hilbert space $\mathcal{H}_{1}\otimes\mathcal{H}_{2}\otimes...\otimes\mathcal{H}_{N}$ is bi-separable if and only if there exists a bipartition $\alpha|\bar{\alpha}$ such that $|\psi\rangle$ can be decomposed as a tensor product $|\psi_{\alpha|\bar{\alpha}}\rangle=|\psi_{\alpha}\rangle\otimes|\psi_{\bar{\alpha}}\rangle$, where $I\!=\!\{1,2,\cdots,N\}$, $\alpha=\{i_1,...,i_m\}$ is an arbitrary nonempty proper subset of $I$, and $\bar{\alpha}=I\setminus\alpha$ is the complementary set of $\alpha$ in the set $I$. If an $N$-partite pure state is not bi-separable then it is genuinely $N$-partite entangled. Similarly, an $N$-partite mixed state $\varrho$ is bi-separable if and only if it can be
written as a convex combination of bi-separable pure states:
\begin{equation}\label{f111}
\varrho=\sum_{\alpha}\bigg(\sum_{k}p_{k}^{\alpha}|\psi_{\alpha}\rangle\langle\psi_{\alpha}|_k\otimes|\psi_{\bar{\alpha}}\rangle\langle\psi_{\bar{\alpha}}|_k\bigg),
\end{equation}
where each component $k$ is bi-separable (possibly under different partitions), and $\sum_{\alpha}$ is for all possible $2^{N-1}-1$ bipartitions. If an $N$-partite mixed state is not bi-separable then it is genuinely $N$-partite entangled.

\begin{figure*}[htb]
\includegraphics[scale=0.67]{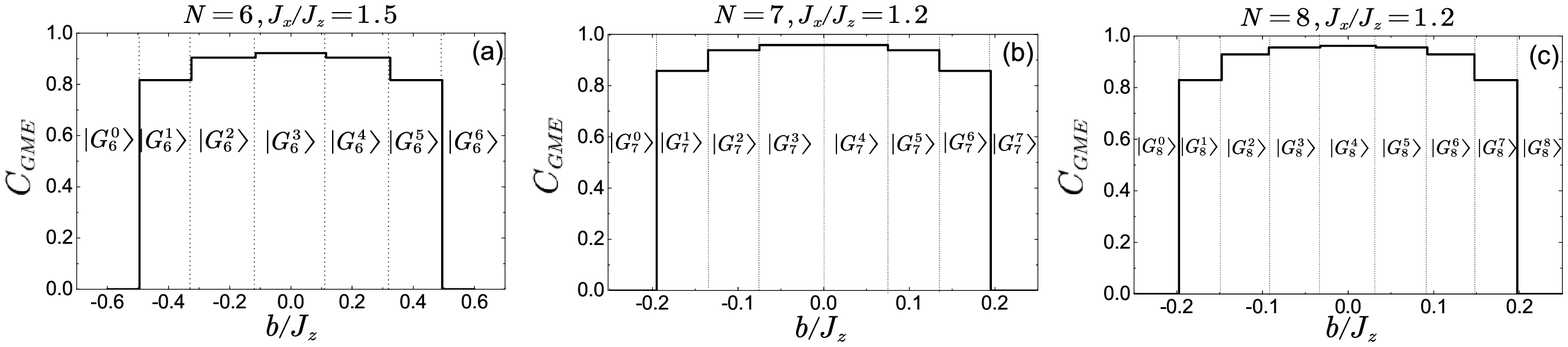}
\caption{The solid line indicates GME concurrence with changing $b/J_z$ and remaining $J_{x}/J_z$ unchanged.  The value of $C_{G\!M\!E}$ of the $6,7,8$-partite spin chain are respectively showed in (a), (b) and (c). $C_{G\!M\!E}$ has a step-like plateaus and a reflection symmetry structure like Fig.\ref{fig1}. Different storey denotes different $|G_{N}^{k}\rangle\ $ in subfigures. From left to right, the GS is $|G_{N}^{0}\rangle,|G_{N}^{1}\rangle,\cdots,|G_{N}^{N}\rangle$, respectively.} \label{fig4}
\end{figure*}

Let us compare GME with GMN.
If the same measurements $M_I$ in Eq. (\ref{f2}) are performed on the $N$-partite mixed bi-separable state $\varrho$, the joint probability distribution with measuring observable $M_i$ with outcome $r_i$ is
\begin{equation}\label{f111}
\tr (\varrho M_{I}^{r_{I}})=\sum_{\alpha}\bigg(\sum_{k}p_{k}^{\alpha}\langle\psi_{\alpha}|M_{\alpha}^{r_{\alpha}}|\psi_{\alpha}\rangle_k  \langle\psi_{\bar{\alpha}}|M_{\bar{\alpha}}^{r_{\bar{\alpha}}}|\psi_{\bar{\alpha}}\rangle_k\bigg),
\end{equation}
where $M_i^{r_i}$ is the project measurement from $M_i$ with outcome $r_i$.
A state satisfies Eq. (\ref{f111}) also satisfies Eq. (\ref{f2}), but the converse of this statement may not be true, since $\langle\psi_{\alpha}|M_{\alpha}^{r_{\alpha}}|\psi_{\alpha}\rangle_k$ in Eq. (\ref{f111}) can be viewed as $P_{\alpha}(r_{\alpha}|M_{\alpha},\lambda)$ in Eq. (\ref{f2}), but conversely $P_{\alpha}(r_{\alpha}|M_{\alpha},\lambda)$ may not be from a quantum state $|\psi_{\alpha}\rangle_k$. Therefore, the set consisting of all bi-separable states is a subset of the set consisting of all bi-local states. That is to say, the set consisting of all GMN states  is the subset of the set consisting of all GME states.

Recently, a notion of generalized concurrence called GME concurrence \cite{CGME} was introduced in the attempt of distinguishing between GME and partial entanglement.
Given an $N$-partite pure state $|\psi\rangle$, let $\alpha=\{j_{1},j_{2},...,j_{i}\}\subset\{1,2,...,N\}$ be a subset inducing a bipartition $j_{1},j_{2},...,j_{i}|j_{i+1},...,j_{N}$. If $C_{\alpha}^{2}(|\psi\rangle):=1-\tr(\varrho_{\alpha}^{2})$, where $\varrho_{\alpha}$ is the reduced density matrix of the subsystems $\alpha$, the GME concurrence (of a pure state) is
\begin{equation}\label{f12}
C_{G\!M\!E}(|\psi\rangle):=\sqrt{\min_{\alpha}C_{\alpha}^{2}(|\psi\rangle)}.
\end{equation}
The state $|\psi\rangle$ is genuinely multipartite entangled if and only if $C_{G\!M\!E}(|\psi\rangle)>0$,

For example, we consider the GSs of a $6$-partite spin chain $|G_{6}^{k}\rangle$ $(k\!=\!1,2,...,\!6)$. In this case, the subset $\alpha$ can contain at most $3$ particles. Then $\alpha$ has a total of 31 combinations: $\{1\},\{2\},...,\{1,2\},\{1,3\},...,\{1,2,3\},\{1,2,4\},...$. Its GME concurrence is then
\begin{equation}\label{f13}
C_{G\!M\!E}^{2}(|G_{6}^{k}\rangle):=\min_{\alpha}\{1-\tr(\rho_{\alpha}^{2})\}.
\end{equation}

\begin{figure*}[htb]
\includegraphics[scale=0.67]{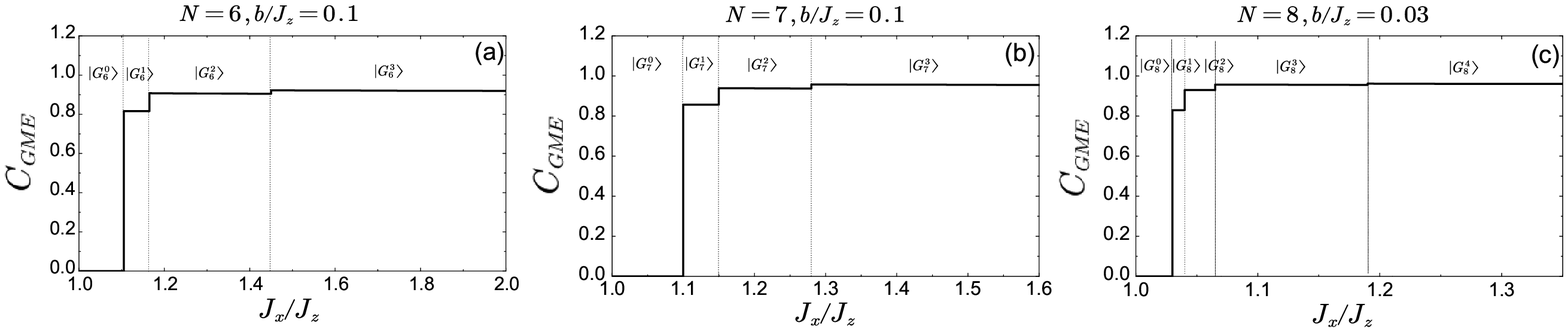}
\caption{The solid line indicates GME concurrence with changing $J_x/J_z$ and remaining $b/J_z$ unchanged.  The value of $C_{G\!M\!E}$ of the $6,7,8$-partite spin chain are respectively showed in (a), (b) and (c). In each subfigure, the GS is $|G_{N}^{0}\rangle,|G_{N}^{1}\rangle,\cdots,|G_{N}^{\lfloor N/2 \rfloor}\rangle$ respectively from left to right. $C_{G\!M\!E}$ varies gradiently at boundary values and decreases tardily with increasing $J_x/J_z$ in each subarea.} \label{fig5}
\end{figure*}

GME concurrence of the $6$,$7$,$8$-partite spin chain is calculated in same conditions as the discussion of GMN. Firstly, We change $b/J_{z}$ and remain $J_{x}/J_{z}$ unchanged. Then we calculate $C_{G\!M\!E}$ in the opposite condition. The results are showed in Fig. \ref{fig4} and Fig. \ref{fig5} respectively.

Comparison results between Fig. \ref{fig1} (Fig. \ref{fig2}) and Fig. \ref{fig4} (Fig. \ref{fig5}) show that where there has GMN, there has GME. It confirms  that the set of all GMN states is a subset of the set of all GME states. The Bell-type inequality is a judgement on GMN, if the result is greater than zero, in other words, violates to the Bell-type inequality, the state must be genuinely multipartite nonlocal. The value of violations is larger at $|G_{N}^{1}\rangle$ and $|G_{N}^{N-1}\rangle$ only because the chosen Bell-type inequality is more sensitive to $|G_{N}^{1}\rangle$ and $|G_{N}^{N-1}\rangle$. Different Bell-type inequalities will cause different results.

Note that in Fig. \ref{fig2}, the value of $J_{x}/J_z$ has a upper bound. When $J_{x}/J_z$ is greater than the upper bound, the violation vanishes but GME still exists. Actually, $C_{G\!M\!E}$ is always possible to be computed with a result greater than zero no matter how large $J_{x}/J_z$ is. There are two possible reasons: (1) The state is not genuinely multipartite nonlocal when $J_{x}/J_z$ is greater than the upper bound. (2) The state is genuinely multipartite nonlocal when $J_{x}/J_z$ is greater than the upper bound, but the chosen Bell-type inequality cannot get violations when $J_{x}/J_z$ is greater than the upper bound and other Bell-type inequalities are needed.

The set of all GMN states is a subset of the set of all GME states. It is pointed out that GMN and GME are equivalent for all entangled symmetric $n$-qubit pure states \cite{ns5}. However, it is also found that there is a mixed state which is genuinely multipartite entangled with fully local hidden variable model,  i.e., this mixed state is a GME state but not a GMN state (it is  not even a nonlocal state) \cite{oppo}. So the set consisting of all GMN states is a \textit{genuine} subset of the set consisting of all  GME states for mixed states, but it is still an open problem whether all pure genuine multipartite entangled states are genuine multipartite nonlocal. From our computation, GMN and GME are equivalent in $6$,$7$,$8$-partite spin chains when $J_{x}/J_z$ is less than a upper bound. The case of large $J_{x}/J_z$ is still not understood.

\section{Discussions and conclusion}
%

In this paper, we have reviewed the definition of GMN in the nonsignaling scenario and a multipartite Bell-type inequality for GMN. A group of measurements has been constructed according to the inequality for the one-dimensional $N$-qubit ferromagnetic chain with nearest-neighbor XXZ interaction, and the results of the violation have been calculated. It is found that the multipartite Bell-type inequality for GMN can be easily violated by the GSs of that anisotropic ferromagnetic spin$-1/2$ chain. We have given detailed numerical results of the violation for GMN inequality in the $6$,$7$,$8$-partite ferromagnetic spin chain, and the analytic results of the violation for the GS state $|G_{N}^{1}\rangle$. We also compute GME concurrence of the $6$,$7$,$8$-partite ferromagnetic spin chain to demonstrate the relationship and difference between GMN and GME.

Genuine multipartite entanglement has been found in some spin chain systems \cite{xy,ising}. But to our knowledge, there are few results about finding GMN in spin chain systems. We have exhibited GMN in the spin chain under the nonsignaling scenario. Furthermore, since the set consisting of all GMN states is a subset of the set consisting of all GME states, the violation of  the multipartite Bell-type inequality for GMN will guarantee the existence of GME in the spin chain as well.

Moreover, our method is possible to be verified by experiments: The XXZ Hamiltonian in Eq. (\ref{f5}) can be implemented in many physical systems such as GaAs and InAs quantum dots \cite{sqd1,sqd2}, neutral atoms in the optical lattice \cite{sqd1}, nitrogen vacancy centers \cite{nvc} and coupled superconducting charge qubits array \cite{cqa}. The GS of systems can be obtained near absolute zero temperature. Then we can perform the joint measurement which is described in Eq. (\ref{f8}) on the system and observe the results whether there is violation to the inequality.

\section*{ACKNOWLEDGMENTS}
This work is funded by the National Natural Science Foundation of China (Grants No. 11504253 and 11474211), the Natural Science Foundation of Jiangsu Province of China (Grant No. BK20141190), the Singapore Ministry of Education (partly through the Academic Research Fund Tier 3 MOE2012-T3-1-009), the National Research Foundation, Singapore (Grant No. WBS: R-710-000-008-271), the open funding program from Key Laboratory of Quantum Information, CAS (Grant No. KQI201605), and the startup funding from Soochow University (Grant No. Q410800215).

\end{document}